\documentclass{PoS}

\PoS{PoS(HEP2005)013}

\title{Phenomenology of black hole evaporation with a cosmological constant}

\ShortTitle{Black hole evaporation with a cosmological content}

\author{J. Labb\'e, \speaker{A. Barrau}, J. Grain\\
        Laboratory for Subatomic Physics and Cosmology\\ 
	CNRS-IN2P3, Joseph Fourier University, France\\
        E-mail: \email{Aurelien.Barrau@cern.ch}}

\abstract{In this brief note, we investigate some possible experimental
consequences of the de-Sitter or Anti-de-Sitter background spacetime structure
for $d$-dimensional evaporating black holes. Possible observational signatures in
Large Hadron Collider (LHC) events are considered in the framework of the
Arkani-Hamed-Dimopoulos-Dvali (ADD) braneworld model. Lower bounds
on the value of the bulk cosmological constant required to produce visible
effects are derived thanks to a dynamical Monte-Carlo simulation. This preliminary
study has to be refined by the accurate computation of the greybody factors. It
opens a new way to investigate the structure of non-asymptotically
flat higher-dimensional spacetimes.}

\FullConference{International Europhysics Conference on High Energy Physics\\
		 July 21st - 27th 2005\\
		 Lisboa, Portugal}

\begin{document}

It has recently been shown  (see \cite{kanti_rev} for a review) that the
fundamental structure of the background spacetime modifies the shape of the
Hawking spectrum for evaporating black holes. This has been extensively studied
to determine the number of extra dimensions \cite{kanti1}. The rotation of the black
hole \cite{kanti2} or a modification of the gravitational theory with higher
order curvature corrections \cite{grain} may also change the characteristics of the emitted spectrum.
In this note, we investigate the possible consequences of a $d$-dimensional
cosmological constant. To study this question into the details, the greybody
factors are required. They have been computed for scalar fields in a
Schwarzshild-de-Sitter spacetime \cite{kanti3} (available at this URL :
http://lpsc.in2p3.fr/ams/greybody/) but they are not yet known for
fermions, gauge bosons, and for a negative cosmological constant \cite{kanti4}.
As a first attempt in this direction which aims at giving rough orders of magnitude,
the present work does not take into account the accurate greybody factors and only deals 
with the consequences of the cosmological constant on the temperature.

\begin{figure}[htbp]
  \centering
  \hspace{ \stretch{1}} 
  \includegraphics[height=5 cm, width=0.55\textwidth]{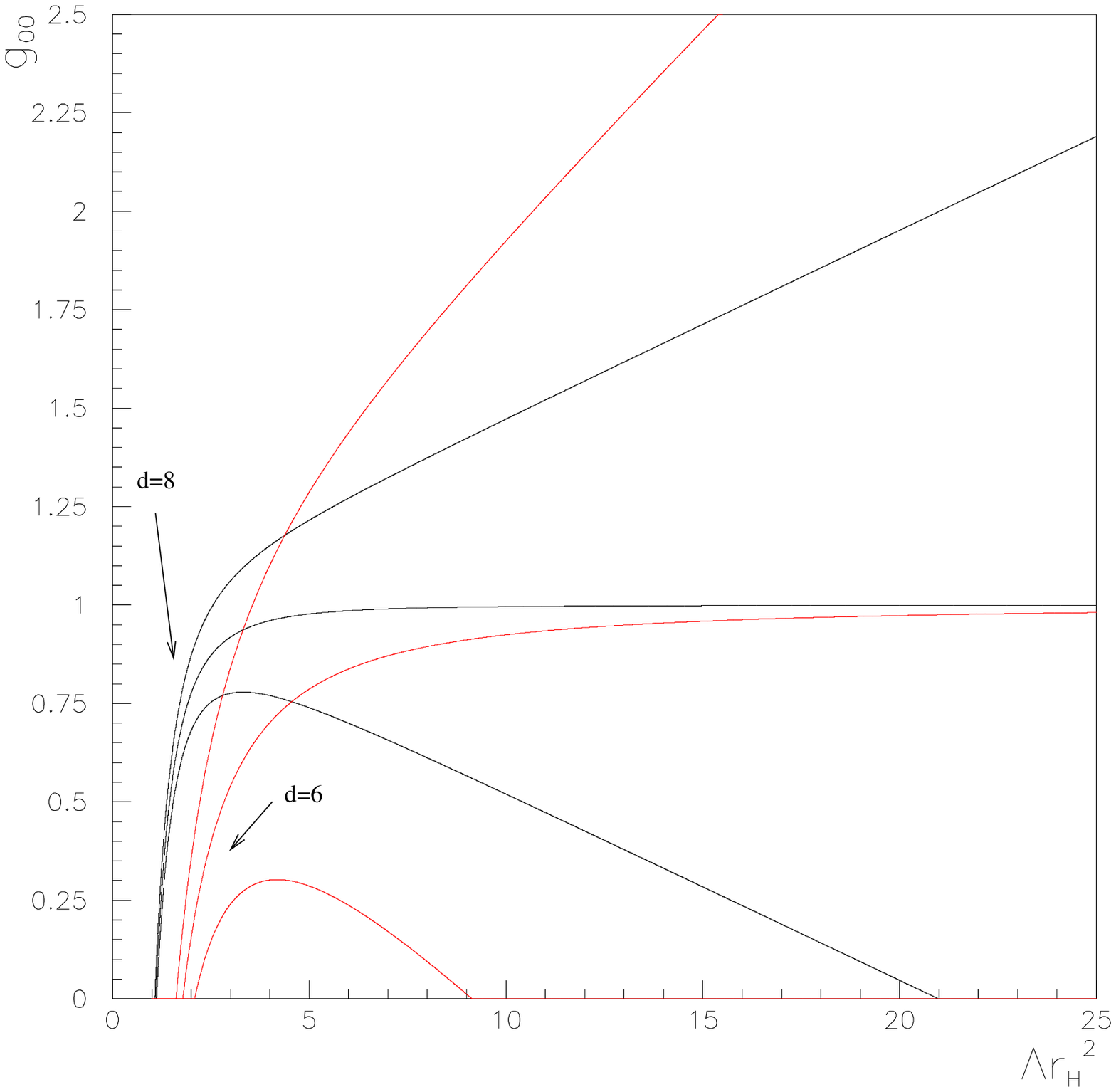}
  \hspace{ \stretch{1}}
  \includegraphics[height=5 cm, width=0.40\textwidth]{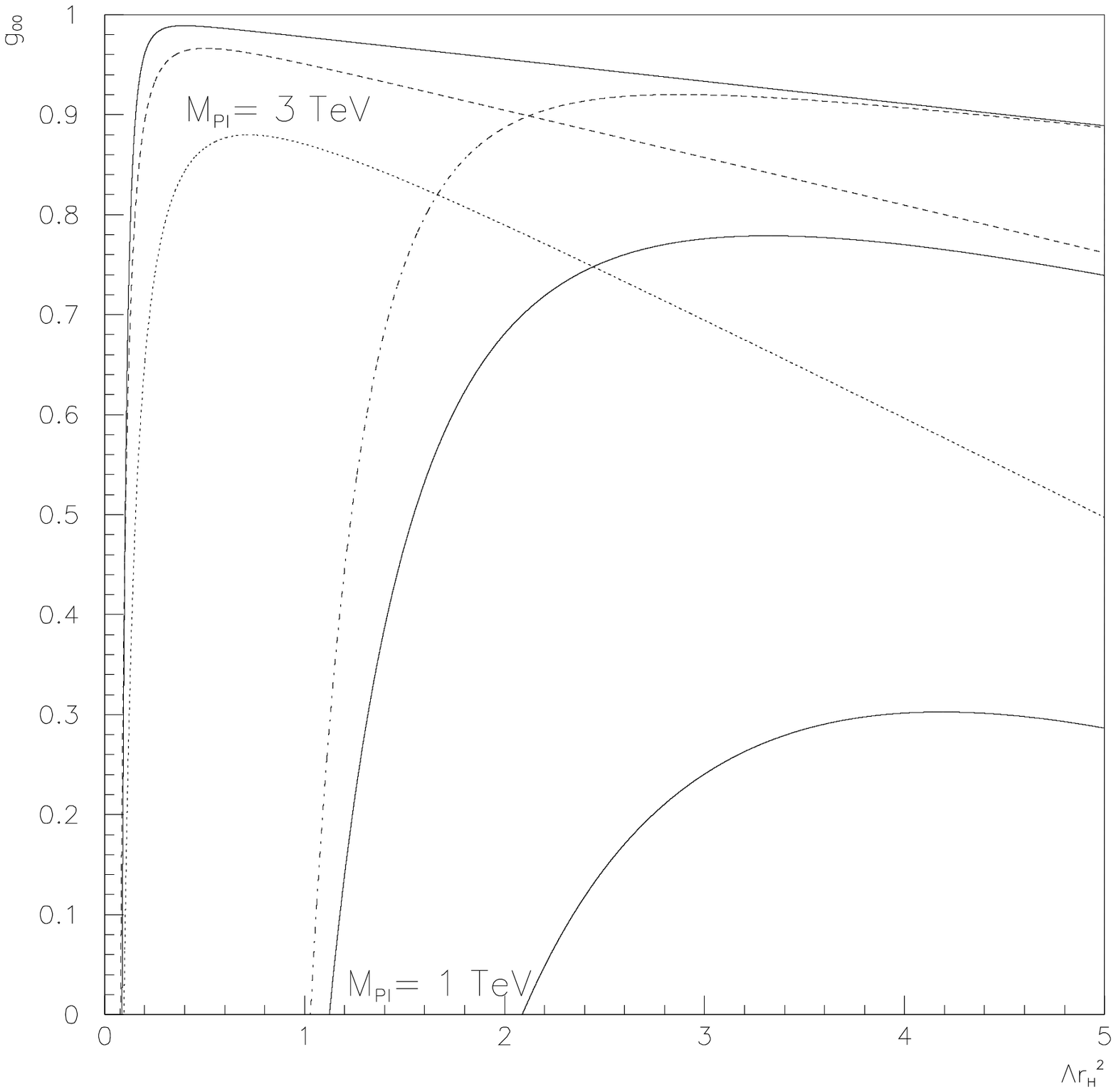}
  \hspace{ \stretch{1}}
  \caption{\small  Metric element $g_{00}=f(r_H)$ for a black hole with mass 
  5~TeV. Left :  $M_{d}=1~\rm{TeV}$ and (from the bottom curve to the upper one)  $\Lambda=1$, 
  $0$ and $-1~\rm{TeV}^2$ for 6 and 8 dimensions. Right : $\Lambda=1~\rm{TeV}^2$
  for two Planck scales : 1 and 3~TeV  and (from the bottom curve to the upper
  one) $d=6,~8,~11$.}
  \label{fig:sds.g00}
\end{figure}

Following \cite{pionniers}, we consider the possible creation of black holes at
colliders in the framework of the ADD model
which allows for a low fundamental Planck scale $M_{d}$ related to the apparent
4-dimensional Planck scale $M_{Pl}$ through  the  volume  of  the
compactified dimensions, $V_{d-4}$, via
$M_{d}=\left(M_{Pl}^2/V_{d-4}\right)^{1/(d-2)}.$
We now consider that the background spacetime is described by a non-zero
cosmological constant $\Lambda$ (which, in $d$ dimension, is not directly constrained by the
tiny value measured by cosmological observations in 4 dimensions). The metric
can be written as 
$ds^2 = g_{\mu \nu}dx^\mu dx^\nu = f(r)dt^2-dr^2/f(r)-r^2 d\Omega_{d-2}^2$
where $d\Omega_{d-2}$ is the solid angle and
$
  g_{00}=f(r)=1 - \frac{16\pi G M}{(d-2)A_{d-2} r^{d-3}} - \frac{2\Lambda r^2}{(d-1)(d-2)}.
$
$G=1/M_{d}^{d-2}$ is the $d$-dimensional gravity constant
and {\small $A_{d-2}=\frac{2\pi^{(d-1)/2}}{\Gamma((d-1)/2)}$} 
is the area of a unit $(d-2)$ sphere. Figure \ref{fig:sds.g00} shows the behavior
of $g_{00}$ for different values of the cosmological constant.
The temperature of the black hole , and the relation between its radius $r_H$ and mass $M$
are given by~:
$$T=\frac{d-3}{4\pi r_H} \left[ 1 - \frac{2 \Lambda r_H^2}{(d-3)(d-2)}
\right]~,~M=\frac{ (d-2)A_{d-2}r_H^{n+1} }{ 16 \pi G } \left[ 1 - \frac{ 2\Lambda r_H^2
}{ (d-3)(d-2) } \right].$$

The consequences of the cosmological constant are more important when the number
of dimensions is smaller. For  $\Lambda>0$, there is a maximum radius
$r_{max}=\sqrt{(d-3)(d-2)/2\Lambda}$ corresponding to a vanishing temperature
when the horizon of the black hole coincide with the cosmological horizon.

\begin{figure}[tbhp]
\centering
  \hspace{ \stretch{1}} 
  \includegraphics[height=4.5 cm, width=0.47\textwidth]{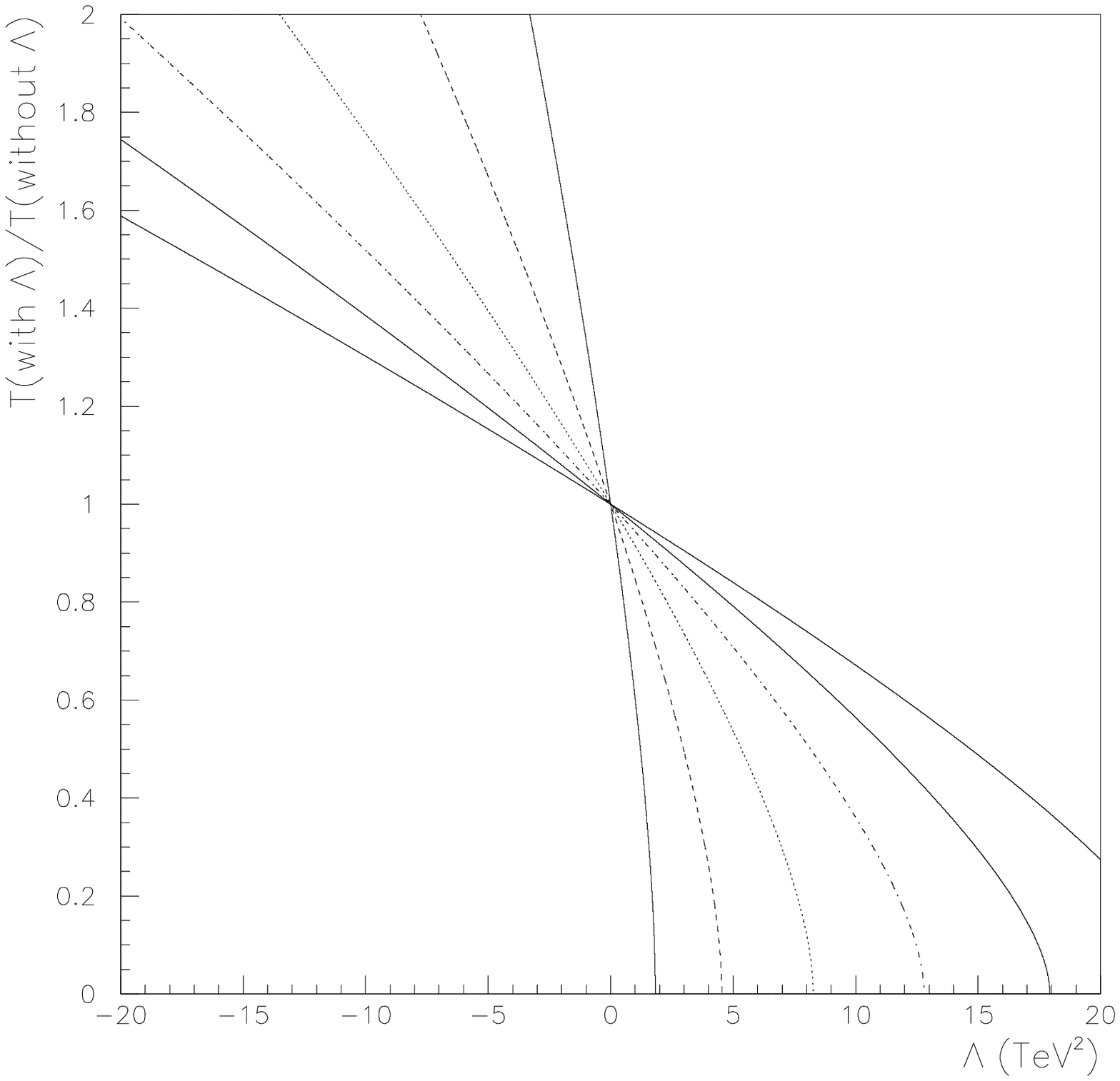}
  \hspace{ \stretch{1}}
  \includegraphics[height=4.5 cm, width=0.47\textwidth]{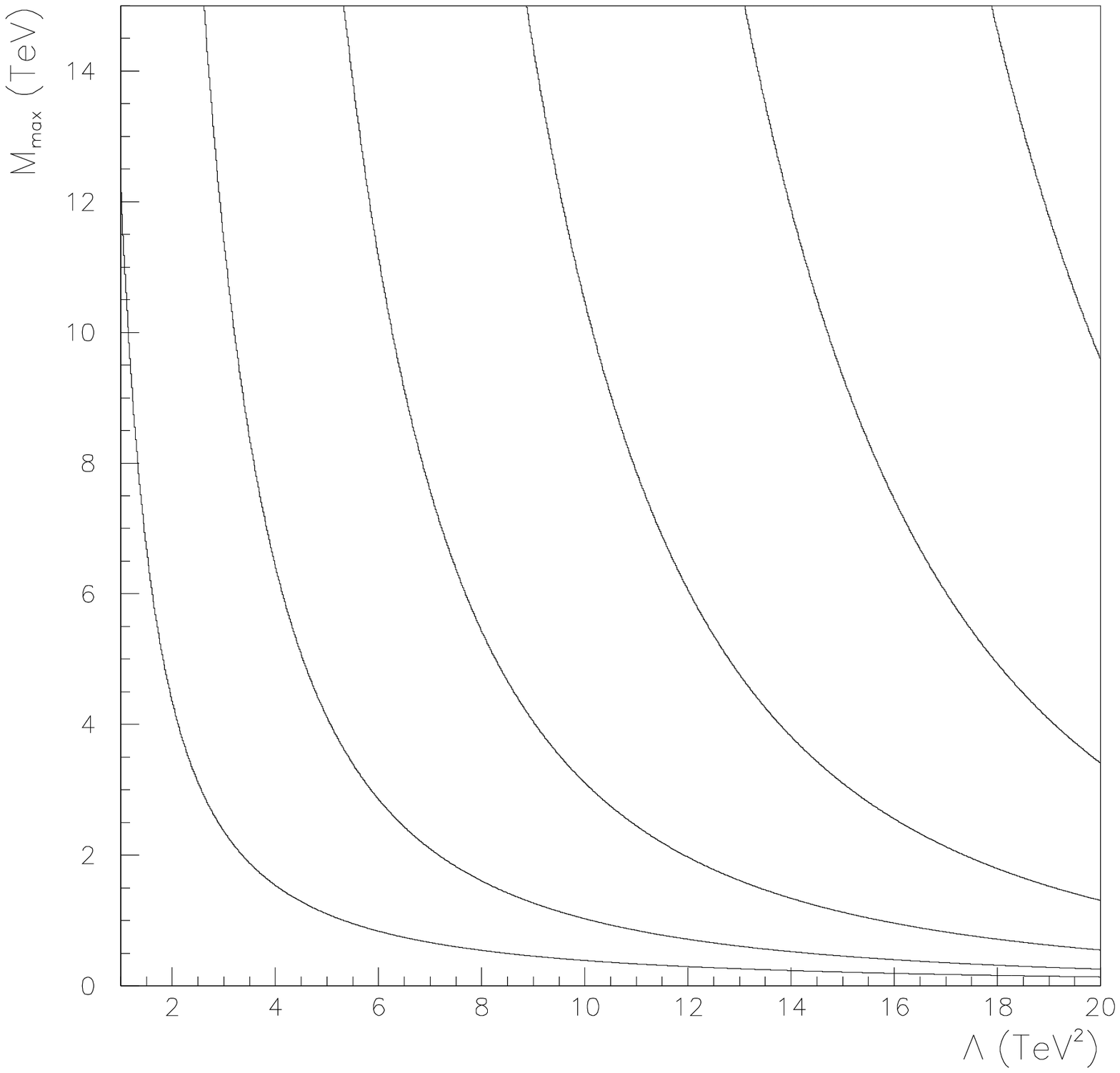}
  \hspace{ \stretch{1}}
  \caption{\small Left : ratio of the temperature with a cosmological constant
  to the temperature without a cosmological constant for a $5~\rm{TeV}$ black
  hole. Right : Maximum mass for $\Lambda=1~\rm{TeV}^2$. Both plots correspond
  to a Planck scale fixed at 1 TeV and to $d$ between 6 and 11.}
  \label{fig:sds.TetMmax}
\end{figure}

With $\Lambda=0$ and $M_d$=1 TeV, the temperature of a black
hole possibly formed at the LHC varies from $0.18$ TeV for $d=6$ to $0.63$ 
TeV for $d=11$ (for $M=5$ TeV). The modifications induced by $\Lambda$, together
with the maximum mass value, are given in Fig. \ref{fig:sds.TetMmax}. It shows
the order of magnitude of a cosmological constant that could be probed at
colliders.

\begin{figure}[tbhp]
\centering
  \hspace{ \stretch{1}} 
  \includegraphics[height=4.5 cm, width=0.55\textwidth]{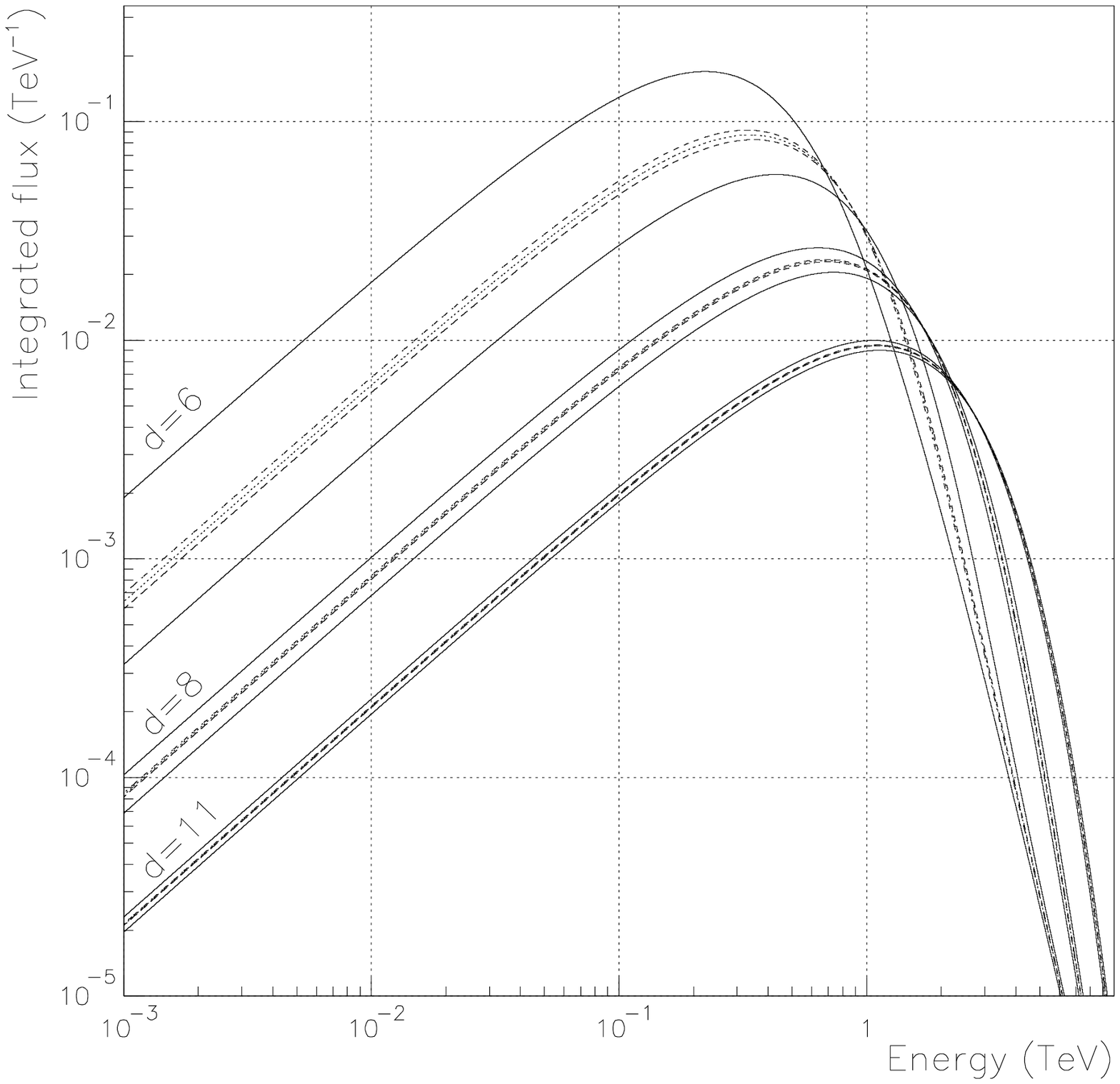}
  \hspace{ \stretch{1}}
  \includegraphics[height=4.5 cm, width=0.40\textwidth]{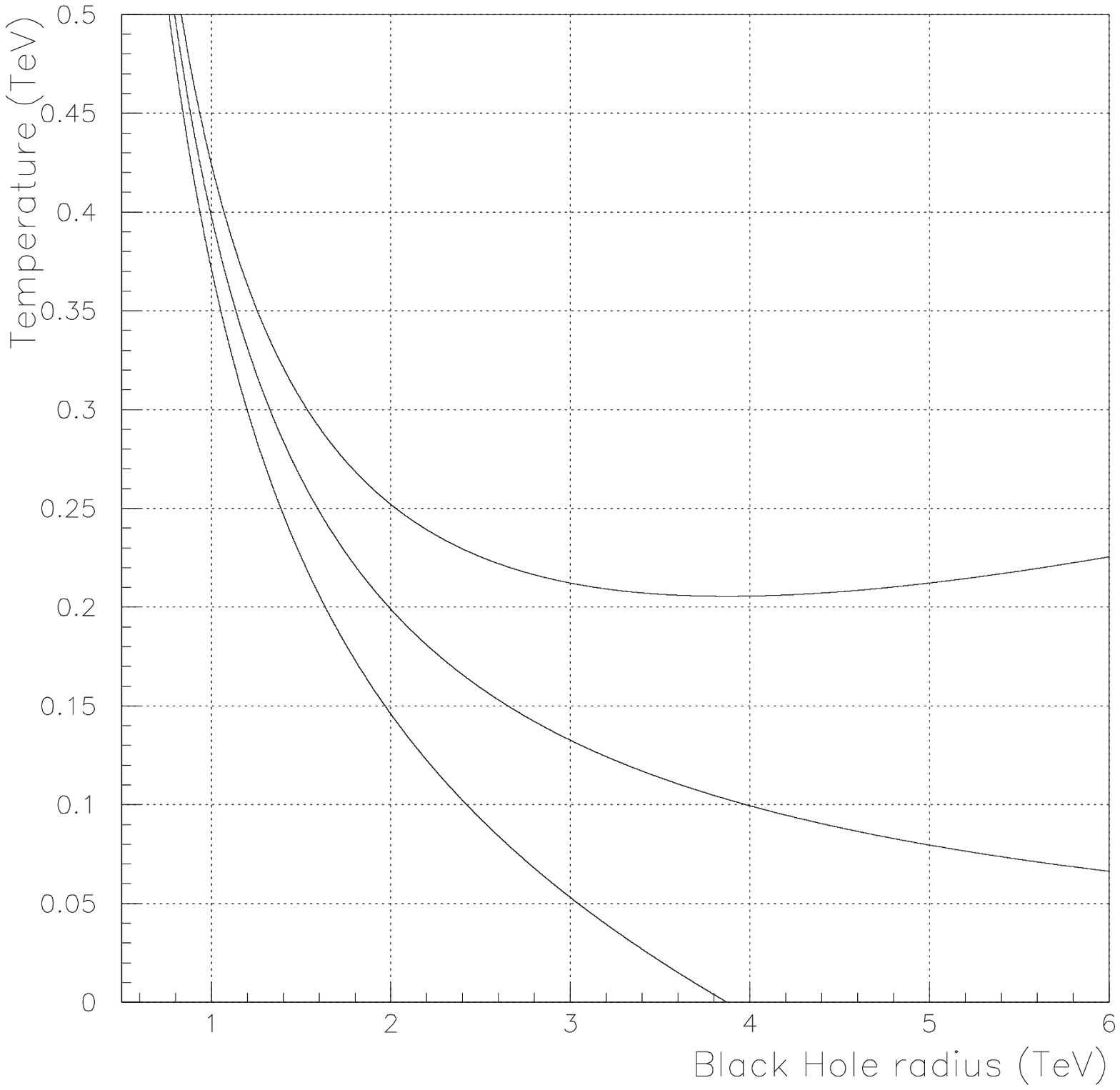}
  \hspace{ \stretch{1}}
  \caption{\small Left : Integrated flux (bosons) for  $M_{d}=1~\rm{TeV}$ and
   $d=\{6,8,11\}$, from top to bottom $\Lambda=\{-1,-0.1,0,0.1,1\}~\rm{TeV}^2$. 
   Right : Temperature of a  $5~\rm{TeV}$ black hole in de-Sitter (bottom), anti-de-Sitter (top)
   and flat spacetimes for $d=8$ and $\vert \Lambda \vert = 1~\rm{TeV}^2$.}
  \label{fig:flux_temp}
\end{figure}

In Fig. \ref{fig:flux_temp} (left), the observable flux is shown, as obtained by integrating over
time the Hawking spectrum with a modified temperature (under the hypothesis of a thermalized
emission). The effects of the cosmological constant are easier to measure with a sudden evaporation
hypothesis, as can be understood from Fig. \ref{fig:flux_temp} (right), as it corresponds to a
larger average radius leading to a highest sensitivity to $\Lambda$ (the curvature induced by the
black-hole being relatively smaller).

Table 1 gives our estimates of the minimum required value of the cosmological
constant to produce measurable effects. It should, of course, be kept in mind that this remains
quite a theoretical approach as the associated size of the horizon is very small.

\begin{table}[t!]
  \label{tab:results}
  \centering
  \caption{\small Minimun cosmological constant that can be seen at the LHC with $100~\rm{fb}^{-1}$.
  In each case : dS, AdS, thermalized emission, instantaneous emission, ideal case
  (upper line), strong
  inelasticity and detector resolution taken into account (lower line). Only the extreme cases (d=6 and d=11
  are displayed).}
  \begin{tabular*}{\textwidth}{p{0.28\textwidth} c c c c c}
    \hline
    \hline
    $\vert \Lambda \vert_{min}$ in $\rm{TeV}^2$
    \rule{0pt}{13pt} % agrandi la hauteur de la première ligne
    &    \multicolumn{2}{c}{Thermalized evaporation}
    &
    &    \multicolumn{2}{c}{Sudden evaporation}\\
    &       dS      &       AdS    &  &       dS      &       AdS\\
    \hline
    $d=6$   &       0.02    &       0.02   & &      0.012   &       0.008\\
    &   0.52    &       0.44   & &      0.30    &       0.22\\[6 pt]
    $d=11$  &       0.18    &       0.18   & &       0.14    &       0.12\\
    &   4.5     &       3.6    & &       3.2     &       2.9\\
    \hline
    \hline
  \end{tabular*}
\end{table}

\begin{figure}[tb]
\centering
  \hspace{ \stretch{1}} 
  \includegraphics[height=5 cm, width=0.47\textwidth]{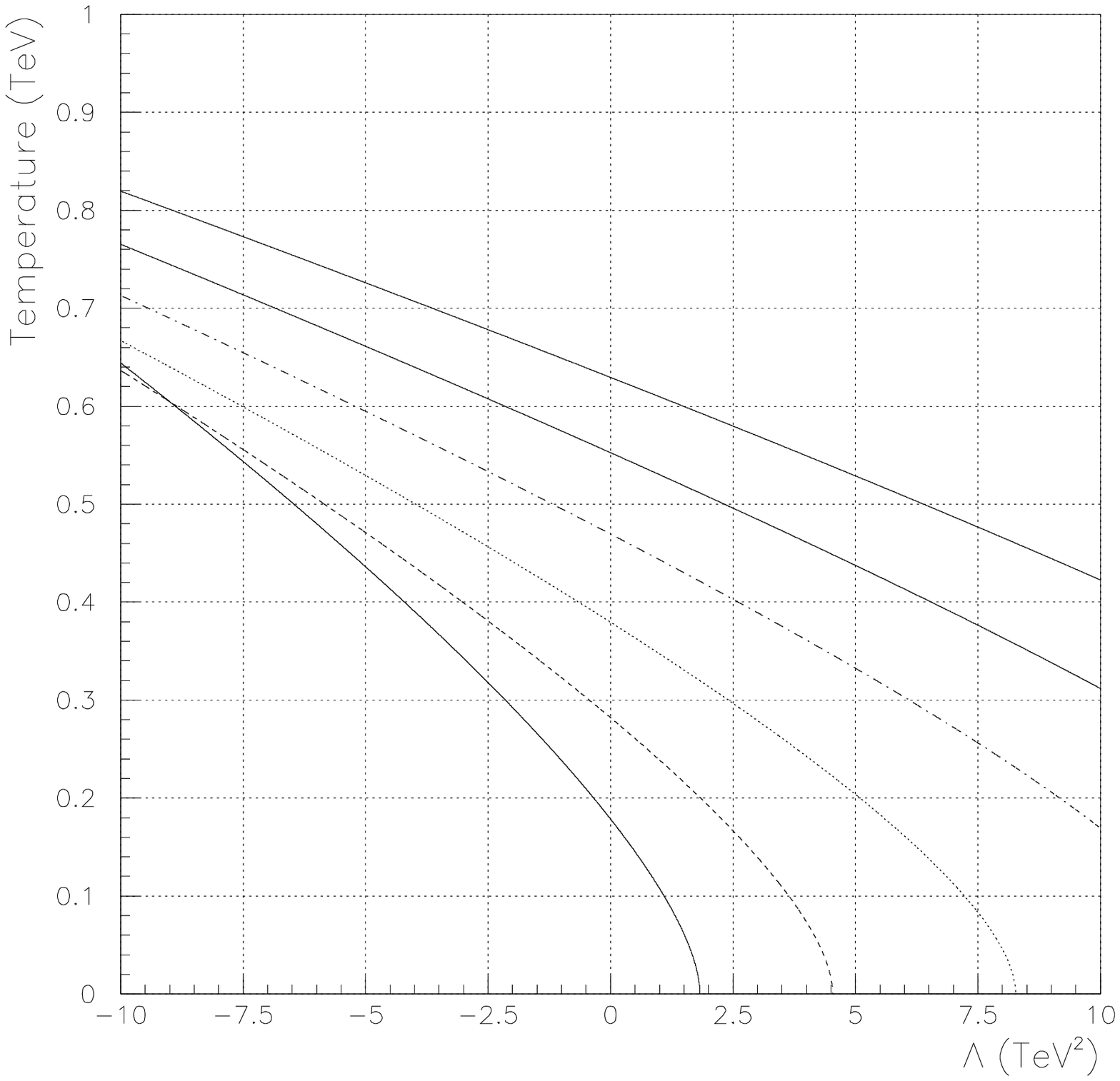}
  \hspace{ \stretch{1}}
  \includegraphics[height=5 cm, width=0.47\textwidth]{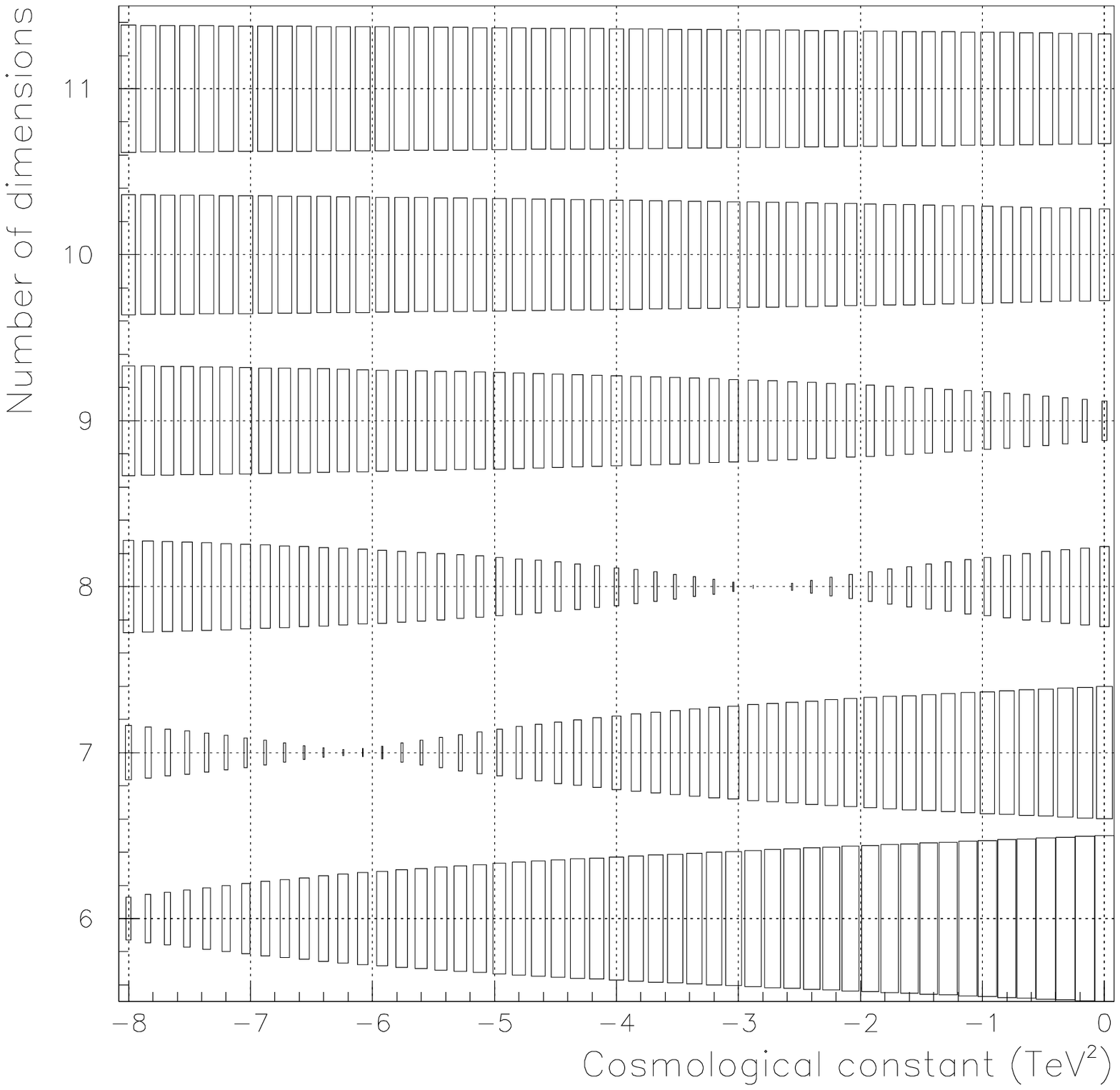}
  \hspace{ \stretch{1}}
  \caption{\small Left : temperature of a $5~\rm{TeV}$ black hole as a function of $\Lambda$ 
  ($d=6$ to $11$ from bottom to top). Right : reduced $\chi^2$ for in put
  parameters : $d=8$ and
  $\Lambda=-2.5~\rm{TeV}^2$.}
  \label{fig:mesure}
\end{figure}

Figure \ref{fig:mesure} shows the results of the Monte-Carlo simulation of the evaporation
process and underlines the degeneracies that could be broken by a detailed study of the greybody
factors. As an example, for input parameters $d=8$ and $\Lambda=-2.5~\rm{TeV}^2$, the statistical
analysis leads to $-4.8  <  \frac{\Lambda}{1~\rm{TeV}^2} < -0.8$ if the dimensionality is assumed
to be otherwise measured. This opens exciting possibilities for experimental probes
of a $d$-dimensional cosmological constant in the bulk.


\begin{thebibliography}{99}
  \bibitem{kanti_rev} P. Kanti, Int. J. Mod. Phys. A 19 (2004) 4899
  \bibitem{kanti1} P. Kanti, I. Olasagasti \& K. Tamvakis, Phys. Rev. D 66
  (2002) 104026;
  P. Kanti \& J. March-Russell, Phys. Rev. D 67 (2003) 104019;
  C.M. Harris \& P. Kanti, JHEP 0310 (2003) 014;
  P. Kanti, J. March-Russell, Phys. Rev. D 66 (2002) 024023
  \bibitem{kanti2} G. Duffy, C. Harris, P. Kanti \& E. Winstanley, hep-th/0507274 
  \bibitem{grain} J. Grain, A. Barrau \& P. Kanti, accepted by Phys. Rev. D,
  hep-th/0509128;
  A. Barrau, J. Grain \& S.O. Alexeyev, Phys. Lett. B 584 (2004) 114
  \bibitem{kanti3}  P. Kanti, J. Grain \& A. Barrau, Phys. Rev. D 71 (2005) 104002
  \bibitem{kanti4}  J. Grain, P. Kanti \& A. Barrau, in preparation
  \bibitem{pionniers}  T. Banks \& W. Fischler, hep-th/9906038;
  S.~Dimopoulos and G.~Landsberg, Phys. Rev. Lett. 87 (2001) 161602;
  S.~B.~Giddings and S.~Thomas, Phys. Rev. D 65 (2002) 056010
  \bibitem{Arkani-Hamed}
  N.~Arkani-Hamed, S.~Dimopoulos \& G.~R.~Dvali,
  Phys.\ Lett.\ B 429 (1998) 263
\end{thebibliography}
\end{document}